\documentclass[10pt,letterpaper]{article}
\usepackage[top=0.85in,left=1.75in,footskip=0.75in,marginparwidth=2.0in]{geometry}

\usepackage[utf8]{inputenc}

\usepackage{cite}

\usepackage{nameref,hyperref}

\usepackage[right]{lineno}

\usepackage{microtype}
\DisableLigatures[f]{encoding = *, family = * }

\usepackage{changepage}

\usepackage[aboveskip=1pt,labelfont=bf,labelsep=period,singlelinecheck=off]{caption}

\makeatletter
\renewcommand{\@biblabel}[1]{\quad#1.}
\makeatother

\usepackage{lastpage,fancyhdr,graphicx}
\usepackage{epstopdf}
\fancyheadoffset[L]{2.25in}
\fancyfootoffset[L]{2.25in}

\usepackage{color}

\definecolor{Gray}{gray}{.25}

\usepackage{graphicx}

\usepackage{sidecap}

\usepackage{wrapfig}
\usepackage[pscoord]{eso-pic}
\usepackage[fulladjust]{marginnote}
\reversemarginpar

\usepackage{enumitem}
\usepackage{float}
\usepackage{multirow}

\usepackage{amsmath,amssymb,amsfonts}
\usepackage{xcolor}
\usepackage{pythonhighlight} 
\usepackage{braket}
\usepackage{subcaption}
\usepackage{rotating}
\usepackage{subcaption}
\usepackage[english]{algorithm2e}
\newcommand*{\defeq}{\stackrel{\text{def}}{=}}

\begin{document}

\vspace*{0.35in}

\begin{flushleft}
	{\Large
		\textbf\newline{Q-Map: Quantum Circuit Implementation of Boolean Functions}
	}
	\newline
	\\
	\textbf{Hassan Hajjdiab}\textsuperscript{1,*},
	\textbf{Ashraf Khalil}\textsuperscript{2},
	\textbf{Hichem Eleuch}\textsuperscript{3,4,5},
	\\
	\bigskip
	$^1$ Computer Science and Software Engineering Department, Concordia University, Montreal, Quebec, Canada;  Email: hassan.hajjdiab@ieee.org
	\bigskip
	
	$^2$ College of Technological Innovation, Zayed University, Abu Dhabi, UAE; Email: ashraf.khalil@zu.ae
	\bigskip
	
	$^3$ Department of applied physics and astronomy,  University of Sharjah, Sharjah, UAE; Email: heleuch@sharjah.ac.ae
	\bigskip
	
	$^4$ College of Arts and Sciences, Abu Dhabi University, Abu Dhabi 59911, UAE
	\bigskip
	
	$^5$ Institute for Quantum Science and Engineering, Texas A\&M University, College Station, TX 77843, USA\\
	\bigskip
	*Author to whom correspondence should be addressed
	
\end{flushleft}

\abstract{Quantum computing has gained attention in  recent  years due to the significant progress in quantum computing technology. Today many companies like IBM, Google and Microsoft   have developed quantum computers and simulators for research and commercial use. The development of quantum techniques and algorithms is essential to exploit the full power of  quantum computers. In this paper we propose a simple visual technique (we call Q-Map)  for quantum realisation of classical  Boolean logic circuits.  The proposed method utilises concepts from Boolean algebra to produce a quantum circuit with minimal number of quantum gates.}



\section*{Introduction}

The advancement  of quantum computing hardware and software intrigues researchers to develop quantum algorithms in  areas such as cryptography, image processing, algorithms, finance   \cite{9172146, app10114040, 9186612,9222275} and many other areas.  One main  advantage of quantum computers compared to classical computers is the processing power. The quantum computer can process  computationally expensive tasks exponentially faster than the classical computer. While classical algorithms are limited in  complexity to  $O(n)$, a quantum search algorithm proposed by Grover \cite{Grover1997} uses  $O(\sqrt{n})$ for unsorted list of $n$ items. Shor \cite{Shor1997}  proposed a quantum algorithm to factor an integer $n$ in polynomial of log $n$ time complexity. At this point, there is no classical algorithm that can solve number factorisation in polynomial time.The  RSA   cryptographic  system \cite{RSA} is based on prime number factorisation, and thus with quantum computers  an RSA encrypted message can be decrypted in polynomial time complexity. Hallgen \cite{Hallgren} presented a polynomial-time quantum algorithm to solve the Pell-Fermat equation \cite{pellEq} ( also known as the Pells equation \footnote{Pell-Fermat equation is: $x^{2}-dy^{2}=1$ and the goal is to find pairs of integers $(x, y)$ to satisfy the equation}). In classical algorithm there is no know polynomial time solution and the problem is know to be NP complete \cite{pellisNP}. Recently, a group of scientists at Google AI Quantum \cite{QuantSuper} used $53$ qubit quantum computer to  sample the output of a pseudo-random quantum circuit \cite{Neill195}.  The results were compared with the sate-of-the art super computer that  needs 10,000 years while the 53 qubit quantum computer needs $200$ seconds.

Another advantage of quantum computers is low  energy consumption. In computation, energy consumption is correlated with reversibility of the computation.   Irreversibly is equivalent to  information erasure. For the case of an \textbf{AND} gate of output  \textbf{0}, we may say that we cannot uniquely identify the input, the \textbf{AND}  operation resulted in erasure of information and thus consumed energy \cite{landauer}. As demonstrated by Landauer \cite{landauer}, classical binary computers mainly dissipate energy during information erasure at the rate of $KT\ln 2$ per bit erased, where $K$ is the  Boltzmann constant and $T$ is the temperature in Kelvins. At room temperature (300 Kelvin), each bit erasure will cost around $3 \times 10^{-21} Jouls$. This number appears to be too small, however the digital binary computer is composed of huge number of Boolean logic gate operations at the hardware level which results in significant consumption of energy.  On the other hand, quantum computers utilise quantum gates which are all reversible gates and thus quantum computation do not result in energy consumption.

In this paper we propose a simple visual technique (we call Q-Map)  for quantum realisation of classical  Boolean logic circuits. Classical boolean computation can be described in terms of classical boolean functions. To perform classical computations using a quantum computer,  the classical boolean function need to be synthesised using reversible functions.

\section*{Overview of classical and Quantum gates}

A Boolean function with $n$ inputs and $m$ outputs is defined as:\\ $\mathcal{B}_{n,m}\defeq \{f|f:\mathbb{B}^{n} \rightarrow \mathbb{B}^{m}\}$, where the \textit{Boolean} values are denoted by $\mathbb{B}\defeq \{0,1\}$.

\begin{figure}[H]

	\includegraphics[width=0.9\linewidth]{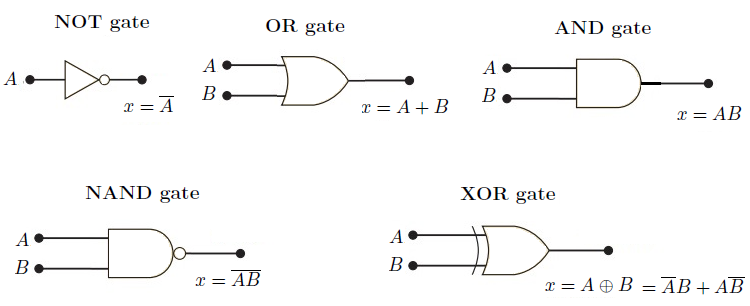}
	\caption{Basic Boolean gates, only the \textbf{NOT} gate is reversible}.
	\label{fig:logicGates}
\end{figure}

 A function $f \in \mathcal{B}_{n,n}$ is reversible if it is bijective \cite{rosen2018}, each input map exactly to one output and the number of inputs is equal to the number of outputs. \\
 Classical Binary computers are, in essence, composed of irreversible logic gates. Other than the \textbf{NOT} gate, the binary logic  gates   are irreversible gates ( see Figure \ref{fig:logicGates}   ). For example the classic \textbf{AND} gate is irreversible, for an output of $\textbf{0}$  the input could be $\textbf{00}$, $\textbf{01}$ or $\textbf{10}$ and  cannot be  uniquely identified.

\begin{figure}[H]
	\centering
	\includegraphics[width=1.0\linewidth]{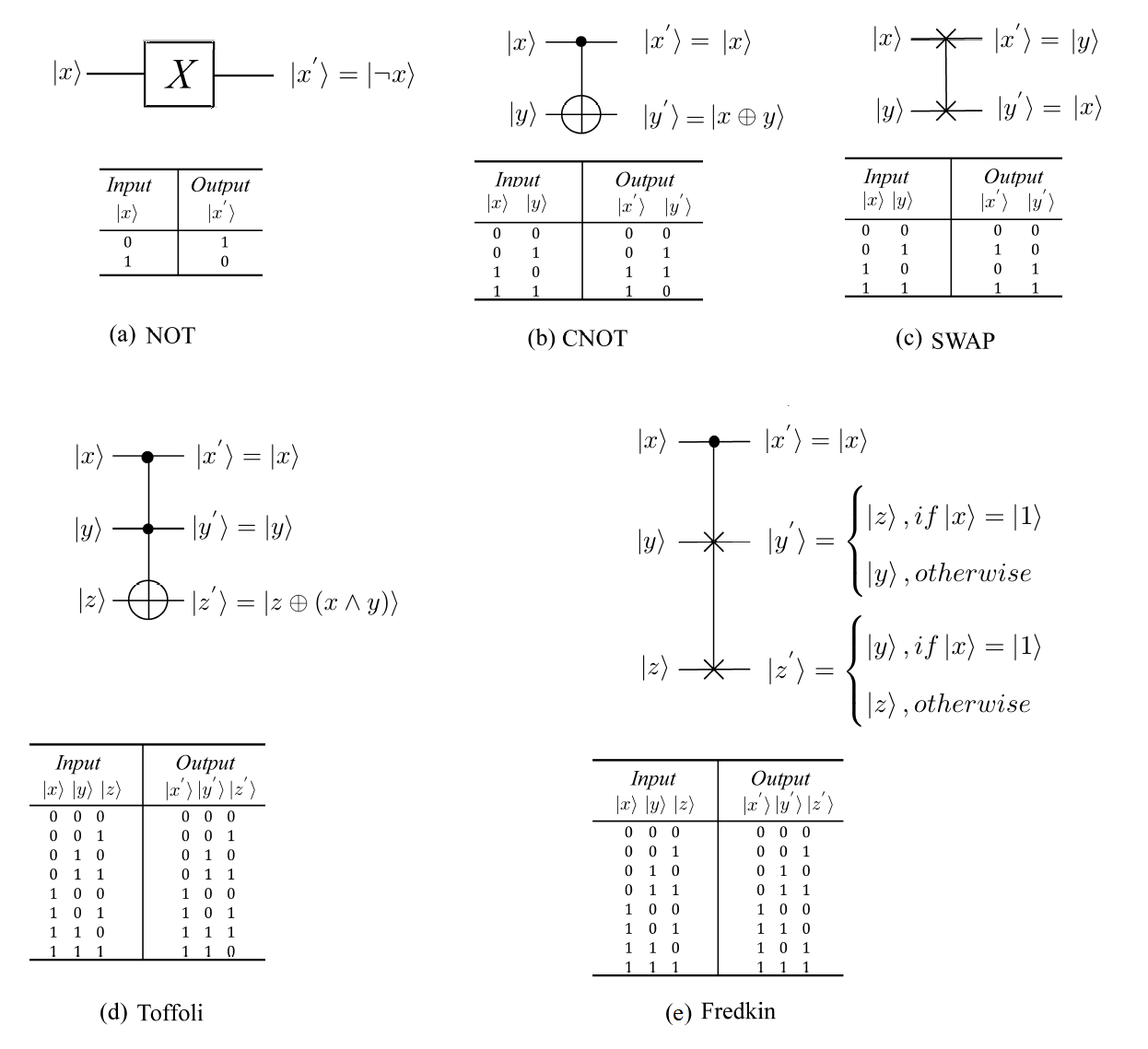}
	\caption{Commonly used quantum gates}
	\label{fig:QGates}
\end{figure}

\begin{figure}[H]
	\centering
	\includegraphics[width=0.9\linewidth]{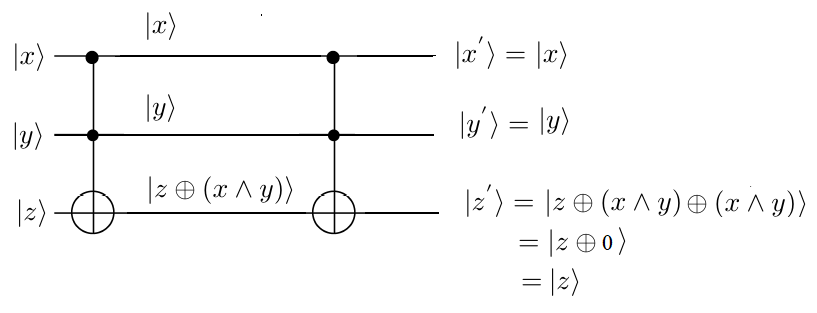}
	\caption{All quantum gates are reversible gates, the input can be recovered by applying the same gate twice. The figure above shows an example using the  Toffoli gate, we can reconstruct the input by applying the Toffolli gate twice.}
	\label{fig:revQgate}
\end{figure}

On the other hand, a quantum bit (or qubit) $\textbf{x}$  represents a unit of information and can be described in a two dimensional quantum system as follows:

$\ket{x}=\left[\begin{array}{c}c_{0}\\ c_{1}\end{array}\right]$ where 
$\sqrt{{\left|c_{0}\right|}^{2}+{\left|c_{1}\right|}^{2}}=1$
\newline \newline
The quantum states of $\ket{0}$  and $\ket{1}$ are represented by the vectors
$\ket{0}=\left[\begin{array}{c}1\\ 0\end{array}\right]$
and  
$\ket{1}=\left[\begin{array}{c}0\\ 1\end{array}\right]$ 
\newline \newline
The qubit can be in an "on" or "off" states as in the classical Boolean computers or in any combination of the "on-off" states : $\textbf{x}=\alpha \ket{0} + \beta \ket{1}$ , where $\sqrt{\alpha^{2}+\beta^2}=1$. To represent a classical digital system the qubit $\ket{0}$ and $\ket{1}$ are sufficient. Figure \ref{fig:QGates} shows the commonly used quantum gates, all quantum gates are reversible and the input can be reconstructed from the output by applying the gate twice. Figure \ref{fig:revQgate} shows an example using the  Toffoli gate, the input can be reconstructed by applying the Toffolli gate twice.\\

\begin{figure}[H]
	\includegraphics[width=1.0\linewidth]{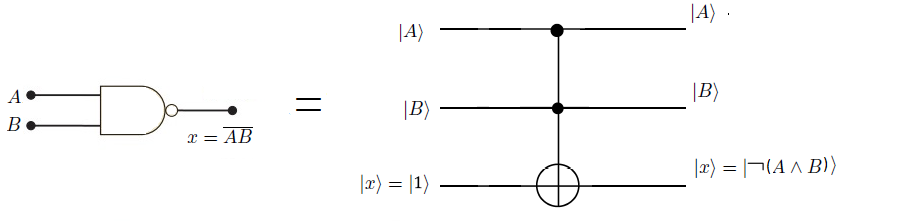}
	\caption{Reconstruction of the logic \textbf{NAND} gate using the Toffolli gate, all Boolean logic gates can be reconstructed using quantum gates.}
	\label{fig:quantumNand}
\end{figure}

In this paper we propose a technique to implement a binary logic circuit using a quantum gates, and thus only the $\ket{0}$ and $\ket{1}$ states of the qubit are utilised. In classical Boolean circuits, the \textbf{NAND} gate is a universal gate and all other Boolean gates could be constructed using one or more \textbf{NAND} gates\cite{Tocci}. 
Any Boolean logic circuit can be designed using reversible quantum gates, the logic gates presented in Figure \ref{fig:logicGates} can  be constructed using a combination of the \textbf{NOT}, \textbf{CNOT} and \textbf{Toffoli} gates, thus the \textbf{NOT-CNOT-Tofolli} gates form a universal basis for quantum circuit implementation \cite{nielsen_chuang_2010,yanofsky,conservativeLogic}. Figure \ref{fig:quantumNand} shows quantum reconstruction of the \textbf{NAND} gate using a Toffoli  gate, the quantum equivalent requires an extra bit ( i.e ancillary bit ).The rest of the boolean gates can be synthesised using the \textbf{NOT-CNOT-Tofolli} bases. 

Thus any Boolean function can be synthesised  by simply replacing each Boolean gate by its quantum counterpart. However this approach is hardly efficient and leads to a significant number of ancillary bits \cite{nabila2016, colin2011}. In literature  several approaches have been proposed \cite{Kerntopf2004, Fazel2007,Soeken2012,SOEKEN20161,Wille2009,devos2008}  to synthesise a given Boolean function with minimal number of ancillary bits. Most of the approaches rely on heuristic methods to minimise the  costs of the resulting circuits  using complex function manipulation \cite{james2006}.  In this paper we present an exact method to realise the quantum implementation of any classical binary system.   The technique (we call Q-Map) is analogous to the  Karnaugh Map \cite{Tocci} for classical logic gate minimisation technique. The main contribution of this paper is as follows:
\begin{itemize}
	\item We propose a visual method to synthesise any  Boolean functions without having to resort to complex function decomposition and manipulation. 
	\item We demonstrate the algorithm  by implementing the 4-bit Gray Code Encoder using  QISKIT. \\

\end{itemize}

 \clearpage		
\section* { Quantum-Map technique}

\begin{figure}[h!]
			\begin{subfigure}[b]{0.7\linewidth}
			\includegraphics[width=\linewidth]{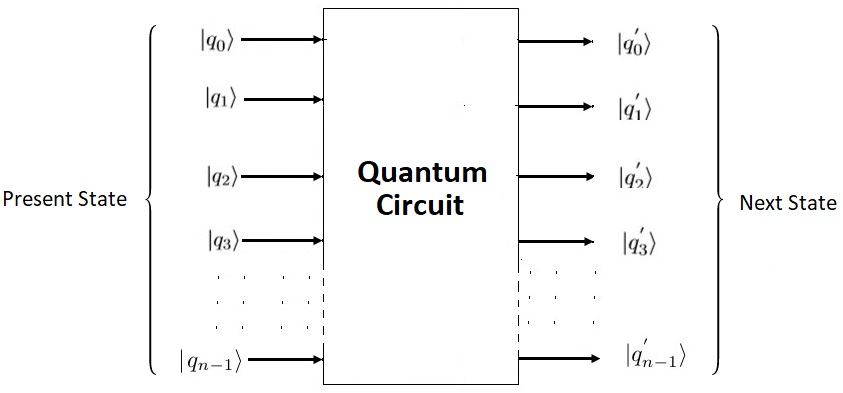}
			\caption{Quantum Circuit}${}_{}$
	\end{subfigure}
	\begin{subfigure}[b]{1.1\linewidth}
		\includegraphics[width=\linewidth]{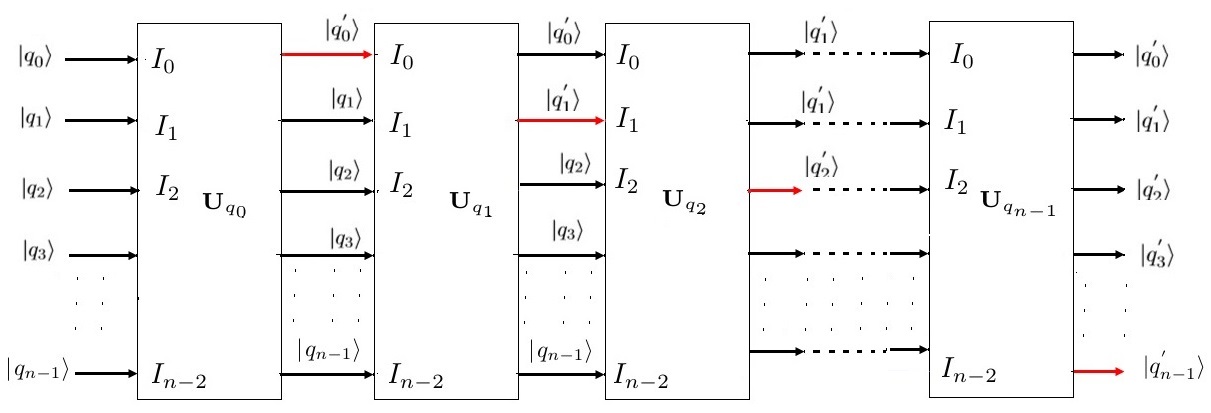}
		\caption{The Quantum circuit is decomposed into $n$ stages.}${}_{}$
	\end{subfigure}
		\caption{(a) The quantum circuit is composed of the present state vector $	\ket {q_{0}q_{1}q_{2}q_{3} \dots q_{n-1}}$ and the next state vector $\ket {q_{0}^{'}q_{1}^{'}q_{3}{'} \dots q_{n-1}^{'}}$ (b) The circuit is decomposed into $n$ cascaded stages, for stage $\textbf{U}_{q_{i}}$ only qubit  $q_{i}$ is changed to its next state  $q_{i}^{'}$ all other qubits are unaltered.}
		\label{fig:QmapStages}
\end{figure}

In our proposed approach, the problem is modelled as a quantum circuit with $n$ input quantum bits ($\ket {q_{0}, q_{1},...,q_{n-1}}$) that represents the initial state of every qubit and $n$ output quantum bits ($\ket{q_{0}^{'}, q_{1}^{'},...,q_{n-1}^{'}}$) that represent the final state of each bit (see	Fig. \ref{fig:QmapStages} (a)). The quantum circuit  is further decomposed into  a series of $n$ cascaded stages . Each stage is  a quantum circuit with $n$ input qubits represented by the vector $\textbf{V}_{i}$  and $n$ output qubits represented by the vector $\textbf{V}_{i}^{'}$  where qubit $\ket {q_{i}}$ is  altered and the rest of the qubits are unaltered, the stages are presented in Fig. \ref{fig:QmapStages} (b).

 \bigskip

\pagebreak

The relationship between the input vector and the output qubit at stage $i$ is defined by a control function $\textbf{U}_{q_{i}}:\mathbb{B}^{n}\longrightarrow \mathbb{B}^{n}$ as follows:
\begin{equation}\label{Eq:Vi}
	\textbf{V}_{i}\stackrel{\textbf{U}_{q_{i} }}{\longrightarrow }\textbf{V}_{i}^{'}
\end{equation}
Where  $\textbf{U}_{q_{i}}$ is a control function that computes a new value at the target output $q_i$ and leaves all other variables unaltered. Equation \ref{Eq:Vi} can be expanded as:
\begin{equation}
	\boxed{
		\begin{array}{rcl}
			\ket {q_{0}q_{1}q_{2}q_{3} \dots q_{n-1}}&\stackrel{\textbf{U}_{q_{0} }}{\longrightarrow }&\ket {q_{0}^{'}q_{1}q_{2}q_{3} \dots q_{n-1}}\\ 
			\ket {q_{0}^{'}q_{1}q_{2}q_{3} \dots q_{n-1}}&\stackrel{\textbf{U}_{q_{1} }}{\longrightarrow }&\ket {q_{0}^{'}q_{1}^{'}q_{2}q_{3} \dots q_{n-1}}\\
			\ket {q_{0}^{'}q_{1}^{'}q_{2}q_{3} \dots q_{n-1}}&\stackrel{\textbf{U}_{q_{2} }}{\longrightarrow }&\ket {q_{0}^{'}q_{1}^{'}q_{2}^{'}q_{3} \dots q_{n-1}}\\
			
			\vdots&\vdots&\vdots\\
			\ket {q_{0}^{'}q_{1}^{'} \dots q_{n-2}^{'}q_{n-1}}&\stackrel{\textbf{U}_{q_{n-1} }}{\longrightarrow }&\ket {q_{0}^{'}q_{1}^{'} \dots q_{n-2}^{'}q_{n-1}^{'}}
		\end{array}
	}
	\label{Eq:array}
\end{equation}

As presented in Equation \ref{Eq:array}, the input vector $\textbf{V}_{0}=\ket {q_{0}q_{1}q_{2}q_{3} \dots q_{n-1}}$ includes all qubits, thus the quantum circuit $\textbf{U}_{q_{0}}$ will calculate $q_{0}^{'}$ (i.e the next state of $q_{0}$) as a function of $n$ qubits.  The output vector $\textbf{V}_{0}^{'}=\ket {q_{0}^{'}q_{1}q_{2}q_{3} \dots q_{n-1}}$ is an $n$ qubit vector where only qubit $q_{0}$ is changed to $q_{0}^{'}$ based on value of the control function $\textbf{U}_{q_{0}}$ and the rest of the qubits are unchanged. To calculate  $q_{1}^{'}$, the output qubit from the previous stage  will be used.   In general, the input vector for any stage $k$ is the vector  $\textbf{V}_{k}=\ket {q_{0}^{'} q_{1}^{'}  \dots q_{k-1}^{'} q_{k} \dots q_{n-1}}$ and the output is the vector $\textbf{V}^{'}_{k}=\ket {q_{0}^{'} q_{1}^{'}  \dots q_{k-1}^{'}q_{k}^{'}  \dots q_{n-1}}$.\\

To calculate $\textbf{U}_{q_{i} }$  for each stage, we follow a function minimisation approach inspired by the Karnaugh Map technique \cite{Tocci} for  Boolean function minimisation. Our proposed approach starts by building a logic map for the function to be minimised we call it the Quantum Map (Q-Map).  The Q-Map is a two-dimensional array of cells used to represent a switching function. The switching function $T(q_{i})$, presented in Table \ref{Table:switchFunction}, represents the toggle state of a qubit from the present state $q_{i}$   to the next state  $q^{'}_{i}$ (i.e toggle from $\ket 0$  to $\ket 1 $  or from $\ket 1$ to  $\ket 0$). The function $T(q_{i})$ can be represented as the logical XOR of the current state  $q_{i}$ with the next state $q_{i}^{'}$ as presented in Table \ref{Table:switchFunction}.

\begin{table}[H]

	\caption{Toggle function to represent change of state} \label{Table:switchFunction}
	
		\begin{tabular}{|p{80pt}|p{80pt}|p{80pt}|}	
		
			\hline
				\centering\multirow{2}{*}{\textbf{\begin{tabular}[c]{@{}c@{}}Present Sate\\ $q_{i}$\end{tabular}}} & 
			
			\centering\multirow{2}{*}{\textbf{\begin{tabular}[c]{@{}c@{}}Next State\\ $q_{i}^{'}$\end{tabular}}} & 
			
		 \multirow{2}{*}{\textbf{\begin{tabular}[c]{@{}c@{}}Toggle function\\ $T(q_{i})=q_{i}\oplus q_{i}^{'}$\end{tabular}}} \\
			&                                                                                            &                                                                                                                      \\ \hline
			
			 $\ket 0$ & 	$\ket  0$   & $\ket 0$                                                                                                               \\ \hline 
			$\ket    0$   & $\ket 1$    & $\ket 1$                                                                                                                   \\ \hline 	
			$\ket     1$      & 	$\ket  
			0$ & $\ket 1$                                                                                                                    \\ \hline 	
			$\ket 1$ & $\ket 1$  & $\ket 0$                                                                                                                    \\ \hline
		\end{tabular}
		
	\end{table}

\bigskip

\begin{figure}[H]
	\centering
	\begin{subfigure}{.4\linewidth}
		\includegraphics[width=\linewidth]{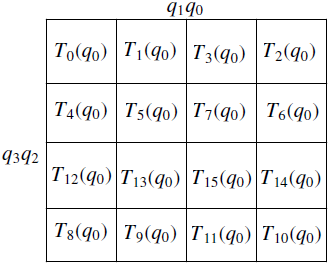}
		\caption{$T(q_{0})$ = $\textstyle f(q_0,q_1,q_2,q_3)$}
		\label{QmapTq0}
	\end{subfigure}\hfill
	\begin{subfigure}{.4\linewidth}
		\includegraphics[width=\linewidth]{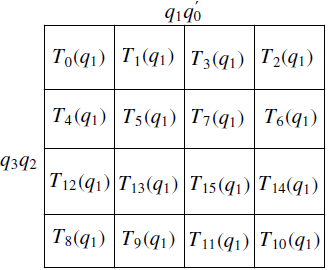}
		\caption{$T(q_{1})$ = $f(q_0^{'},q_1,q_2,q_3)$}
		\label{QmapTq1}
	\end{subfigure}\hfill
	\medskip 
	\begin{subfigure}{.4\linewidth}
		\includegraphics[width=\linewidth]{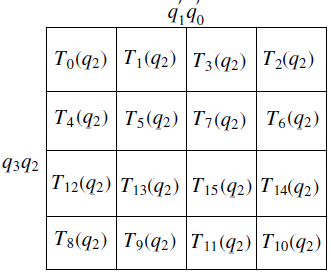}
		\caption{$T(q_{2})$ = $f(q_0^{'},q_1^{'},q_2,q_3)$}
		\label{QmapTq2}
	\end{subfigure}\hfill 
	\begin{subfigure}{.4\linewidth}
		\includegraphics[width=\linewidth]{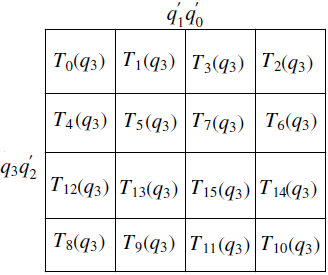}
		\caption{$T(q_{3})$ = $f(q_0^{'},q_1^{'},q_2^{'},q_3)$}
		\label{QmapTq3}
	\end{subfigure}
	\caption{Quantum maps with  four variables. In each cell the corresponding value of $T(q_{i})$  is inscribed.}
	\label{fig:Qmap4Var}
\end{figure}

$T(q_{i})$ is a function of $n$ qubits denoted by  $\ket {q_{0}q_{1}\dots q_{n-1}}$, the Q-Map of the stage $\textbf{U}_{q_{i} }$ is composed of two vectors $\textbf{S}_{1}=\ket{q_{n-1}\dots q_{k}}$ and $\textbf{S}_{2}=\ket{q_{k-1}\dots q_{0}}$ with a row for each assignment of $\textbf{S}_{1}$ for a total of $2^{n-k-1}$ and with a column for each assignment of $\textbf{S}_{2}$ for a total of $2^{k}$ columns. 
Similar to the Karnaugh Map, the adjacency condition is established by labelling the rows and the columns such that for any $2^{r}$ adjacent rows ( or columns) differ only in $r$ variables.
Fig. \ref{fig:Qmap4Var} shows the Q-Map to find the quantum circuit  with four variables. In each cell the corresponding value of the switching function $T(q_{i})$ is inscribed. Using the Q-Map, a minimal expression $T(q_{i})$ is calculated for each qubit. And finally the quantum circuit is implemented using the $\bf{NOT-CNOT-Toffoli}$ quantum gate basis.

To demonstrate our proposed approach,  we present a quantum circuit  implementation of the Gray Code to Binary converter. The details of proposed algorithm with the reversible circuit implementation are presented in  the next section.

\clearpage
\section*{Implementation of the Gray Code Encoder}\label{sec:gray}

In this section we demonstrate the technique by implementing the Gray Code to Binary converter. The Gray Code is used in many applications such as position control systems, communications and many other areas \cite{Todd2020}. The Gray code provides a binary code that changes by one bit only when it changes from one state  to the next. The Gray code and the corresponding decimal unsigned binary  equivalent is shown in Table \ref{table:grayCodeTruthTable}.

\begin{table}[H]
	\centering
	\caption{Gray code to Binary converter truth table.	\label{table:grayCodeTruthTable} }
	
			\begin{tabular}{|p{80pt}|p{80pt}|}
				
				\hline
				\textbf{Present State}       & \textbf{Next State}  \\ 
				$\textbf{q}_{3}\  \textbf{q}_{2}\ \textbf{q}_{1}\ \textbf{q}_{0}$ & ${\textbf{q}_{3}^{'}\ \textbf{q}_{2}^{'}\ \textbf{q}_{1}^{'}\ \textbf{ q}_{0}^{'}}$  \\ 
				\hline
				${0\quad 0\quad 0\quad 0}$   & ${0\quad 0\quad 0\quad 0}$      \\ 
				${0\quad 0\quad 0\quad 1}$      & ${0\quad 0\quad 0\quad 1}$             \\ 
				${0\quad 0\quad 1\quad 1}$        & ${0\quad 0\quad 1\quad 0}$              \\ 
				${0\quad 0\quad 1\quad 0}$                 & ${0\quad 0\quad 1\quad 1}$                 \\ 
				${0\quad 1\quad 1\quad 0}$                 & ${0\quad 1\quad 0\quad 0}$               \\ 
				${0\quad 1\quad 1\quad 1}$                 & ${0\quad 1\quad 0\quad 1}$           \\ 
				${0\quad 1\quad 0\quad 1}$                 & ${0\quad 1\quad 1\quad 0}$            \\ 
				${0\quad 1\quad 0\quad 0}$                 & ${0\quad 1\quad 1\quad 1}$           \\ 
				${1\quad 1\quad 0\quad 0}$                 & ${1\quad 0\quad 0\quad 0}$             \\ 
				${1\quad 1\quad 0\quad 1}$                 & ${1\quad 0\quad 0\quad 1}$               \\ 
				${1\quad 1\quad 1\quad 1}$                 & ${1\quad 0\quad 1\quad 0}$                    \\ 
				${1\quad 1\quad 1\quad 0}$                 & ${1\quad 0\quad 1\quad 1}$                 \\ 
				${1\quad 0\quad 1\quad 0}$                 & ${1\quad 1\quad 0\quad 0}$                   \\ 
				${1\quad 0\quad 1\quad 1}$                 & ${1\quad 1\quad 0\quad 1}$                   \\ 
				${1\quad 0\quad 0\quad 1}$                 & ${1\quad 1\quad 1\quad 0}$               \\ 
				${1\quad 0\quad 0\quad 0}$                 & ${1\quad 1\quad 1\quad 1}$                 \\ 
				\hline
			\end{tabular}
					
		\end{table}

The first step starts by building  the switching function $T(q_{i})$ for each quantum bit as described in Table \ref{Table:switchFunction}. The function $T(q_{i})$ for each qubit is calculated as:
\begin{eqnarray}\label{eq:FQ}
	T(q_{0}) =  q_{0} \oplus q_{0}^{'}  \\
	T(q_{1}) =  q_{1} \oplus q_{1}^{'}  \\
	T(q_{2}) =  q_{2} \oplus q_{2}^{'}  \\
	T(q_{3}) =  q_{3} \oplus q_{3}^{'}  
\end{eqnarray}

The result is presented in Table \ref{table:grayCodeSwitchtable}

\begin{table}[H]
	\centering
	\caption{Gray Code to binary converter: Q-Map functions are calculated as the XOR of the present and next states.	\label{table:grayCodeSwitchtable} }

		\begin{tabular}{|p{80pt}p{80pt}|p{18pt}p{18pt}p{18pt}p{18pt}|}
		
			\hline
		
			\textbf{Present State} & \textbf{Next State} & \multicolumn{4}{c|}{\textbf{Toggle Functions}} \\ 
			$\textbf{q}_{3}\  \textbf{q}_{2}\ \textbf{q}_{1}\ \textbf{q}_{0}$ & ${\textbf{q}_{3}^{'}\ \textbf{q}_{2}^{'}\ \textbf{q}_{1}^{'}\ \textbf{ q}_{0}^{'}}$ & $\textbf{T}({\textbf{q}_{3}})$ & $\textbf{T}({\textbf{q}_{2}})$ & $\textbf{T}({\textbf{q}_{1}})$ & $\textbf{T}({\textbf{q}_{0}})$ \\ 
		
			\hline
			${0\quad 0\quad 0\quad 0}$                 & ${0\quad 0\quad 0\quad 0}$                                 & ${0}$        & ${0}$        & ${0}$        & ${0}$        \\ 
			${0\quad 0\quad 0\quad 1}$                 & ${0\quad 0\quad 0\quad 1}$                                 & ${0}$        & ${0}$        & ${0}$        & ${0}$        \\ 
			${0\quad 0\quad 1\quad 1}$                 & ${0\quad 0\quad 1\quad 0}$                                 & ${0}$        & ${0}$        & ${0}$        & ${1}$        \\ 
			${0\quad 0\quad 1\quad 0}$                 & ${0\quad 0\quad 1\quad 1}$                                 & ${0}$        & ${0}$        & ${0}$        & ${1}$        \\ 
			${0\quad 1\quad 1\quad 0}$                 & ${0\quad 1\quad 0\quad 0}$                                 & ${0}$        & ${0}$        & ${1}$        & ${0}$        \\ 
			${0\quad 1\quad 1\quad 1}$                 & ${0\quad 1\quad 0\quad 1}$                                 & ${0}$        & ${0}$        & ${1}$        & ${0}$        \\ 
			${0\quad 1\quad 0\quad 1}$                 & ${0\quad 1\quad 1\quad 0}$                                 & ${0}$        & ${0}$        & ${1}$        & ${1}$        \\ 
			${0\quad 1\quad 0\quad 0}$                 & ${0\quad 1\quad 1\quad 1}$                                 & ${0}$        & ${0}$        & ${1}$        & ${1}$        \\ 
			${1\quad 1\quad 0\quad 0}$                 & ${1\quad 0\quad 0\quad 0}$                                 & ${0}$        & ${1}$        & ${0}$        & ${0}$        \\ 
			${1\quad 1\quad 0\quad 1}$                 & ${1\quad 0\quad 0\quad 1}$                                 & ${0}$        & ${1}$        & ${0}$        & ${0}$        \\ 
			${1\quad 1\quad 1\quad 1}$                 & ${1\quad 0\quad 1\quad 0}$                                 & ${0}$        & ${1}$        & ${0}$        & ${1}$        \\ 
			${1\quad 1\quad 1\quad 0}$                 & ${1\quad 0\quad 1\quad 1}$                                 & ${0}$        & ${1}$        & ${0}$        & ${1}$        \\ 
			${1\quad 0\quad 1\quad 0}$                 & ${1\quad 1\quad 0\quad 0}$                                 & ${0}$        & ${1}$        & ${1}$        & ${0}$        \\ 
			${1\quad 0\quad 1\quad 1}$                 & ${1\quad 1\quad 0\quad 1}$                                 & ${0}$        & ${1}$        & ${1}$        & ${0}$        \\ 
			${1\quad 0\quad 0\quad 1}$                 & ${1\quad 1\quad 1\quad 0}$                                 & ${0}$        & ${1}$        & ${1}$        & ${1}$        \\ 
			${1\quad 0\quad 0\quad 0}$                 & ${1\quad 1\quad 1\quad 1}$                                 & ${0}$        & ${1}$        & ${1}$        & ${1}$        \\ 
		
			\hline
		\end{tabular}

	\end{table}

The second step is to establish the Q-Map for each qubit.   Figure \ref{fig:Qmap} (a) shows the Q-Map to evaluate ${q_{0}}$ given the values of the  input ${q_{1}}$, ${q_{2}}$ and $q_{3}$. A value of $1$ in the Q-Map represents the state of ${q_{1}}$, $q_{2}$ and ${q_{3}}$ when ${q_{0}}$ toggles its state. Thus  ${q_{0}}$ will toggle its state when  the following expression is true:

\begin{align}\label{eq:QmapXORq0}
	T(q_{0}) = \overline{q}_{3} \overline{q}_{2} q_{1} \oplus \overline{q}_{3} q_{2} \overline{q}_{1} \oplus q_{3} \overline{q}_{2} \overline{q}_{1} \oplus q_{3}q_{2}q_{1}
\end{align}

Figure \ref{fig:Qmap} (b) shows the Q-Map  to find  $T(q_{1})$. The expression is calculated  based on inputs  $q_{3}$, $q_{2}$ and $q_{0}^{'}$  as follows:
\begin{align}\label{eq:QmapXORq1}
	T(q_{1}) = q_{3} \overline{q}_{2} \oplus \overline{q}_{3} q_{2}
	\end{align}

Figure \ref{fig:Qmap} (c) shows the Q-Map  to find  $T(q_{2})$. The expression is calculated based on  inputs  $q_{3}$, $q_{1}^{'}$ and $q_{0}^{'}$  as follows:
\begin{align}\label{eq:QmapXORq2}
	T(q_{2}) = q_{3}
\end{align}

Finally, figure \ref{fig:Qmap} (d) shows the Q-Map  to find  $T(q_{3})$. The expression is calculated based on the inputs  $q_{2}^{'}$, $q_{1}^{'}$ and $q_{0}^{'}$  as follows:

\begin{figure}[H]
	\centering
	\begin{subfigure}{.33\linewidth}
		\includegraphics[width=0.9\linewidth]{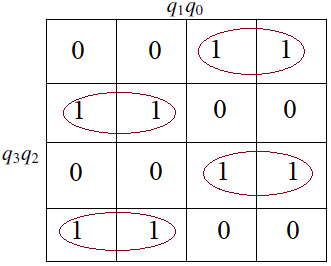}
		\caption{$T(q_{0})=f(q_0,q_1,q_2,q_3)$}
		\label{QmapTq0}
	\end{subfigure}
	\begin{subfigure}{.33\linewidth}
		\includegraphics[width=0.9\linewidth]{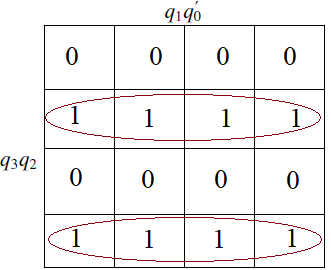}
		\caption{$T(q_{1})=f(q_0^{'},q_1,q_2,q_3)$}
		\label{QmapTq1}
	\end{subfigure}
	\medskip 
	
	\begin{subfigure}{.33\linewidth}
		\includegraphics[width=0.9\linewidth]{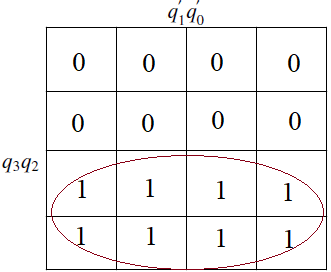}
		\caption{$T(q_{2})=f(q_0^{'},q_1^{'},q_2,q_3)$}
		\label{QmapTq2}
	\end{subfigure}
	\begin{subfigure}{.33\linewidth}
		\includegraphics[width=0.9\linewidth]{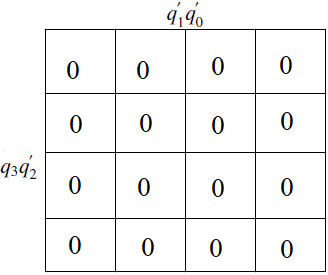}
		\caption{$T(q_{3})=f(q_0^{'},q_1^{'},q_2^{'},q_3)$}
		\label{QmapTq3}
	\end{subfigure}
	\caption{Q-Map representing the switching function for qubits $q_{0}$, $q_{1}$, $q_{2}$ and $q_{3}$,  $T(q_i)$ is written in Exclusive-OR Sum-of-Product form (ESOP)}
	\label{fig:Qmap}
\end{figure}

\medspace

 Notice here we use the XOR ($\oplus$) operation and the function $T(q_i)$ is represented in Exclusive-OR Sum-of-Product form (ESOP) form since the qubit will toggle only if we have an odd number of true terms. For even number of true terms the qubit will retain its initial state.  In classical Boolean function minimisation using Karnaugh Map the logic OR  ($+$) operation is used and the function is expressed in  Sum of Product (SOP) form  since the Boolean function will be true if any of the terms is true and so group overlap is allowed.

In our proposed Q-Map approach group overlap is not allowed, this makes all terms in  every function $T(q_i)$ mutually exclusive and only one term can be true at one instant of time. Thus the  XOR ($\oplus$) operation can be replaced by the OR ($+$) operation and the function $T(q_i)$ can be represented in  the Sum of Product (SOP) form. Since each Q-Map in  Figure \ref{fig:Qmap} contains no overlapping groups, the functions can be represented in Sum of Product (SOP) form as follows:

\begin{equation}\label{eq:T}
	\boxed{
		\begin{array}{rcl}
			T(q_{0}) & = & \overline{q}_{3} \overline{q}_{2} q_{1} + \overline{q}_{3} q_{2} \overline{q}_{1} + q_{3} \overline{q}_{2} \overline{q}_{1} + q_{3}q_{2}q_{1}\\
			T(q_{1}) & = & q_{3} \overline{q}_{2} + \overline{q}_{3} q_{2}\\
			T(q_{2}) & = & q_{3}\\
			T(q_{3}) & = & 0
		\end{array}
	}
\end{equation}

\bigskip

\begin{figure}[h]
	\centering
	\includegraphics[width=1.0\textwidth]{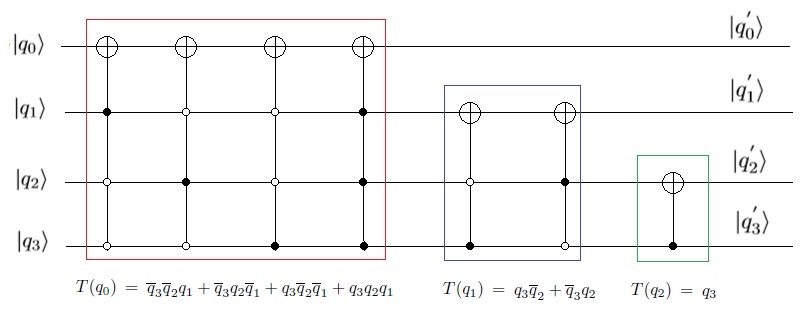}
	\caption{Implementation of the Quantum circuit using 2 and 3 input Toffoli and CNOT gates}
	\label{fig:Qcircuit1}
\end{figure}

The  functions presented  in Equation \ref{eq:T} can be realised by a reversible circuit with only four lines (i.e. $O(n)$ lines) using the \textbf{NOT-CNOT-Toffoli} bases with Multi-Control Toffolli gates \cite{Shende2003}.  This means that the implementation is efficient and no temporary lines (also referred as ancilla)   are needed. The quantum circuit can be implemented using  CNOT and two  and three  input Toffoli gates as shown in Figure \ref{fig:Qcircuit1}.

\begin{figure}[H]
	\centering
	\includegraphics[width=1.0\textwidth]{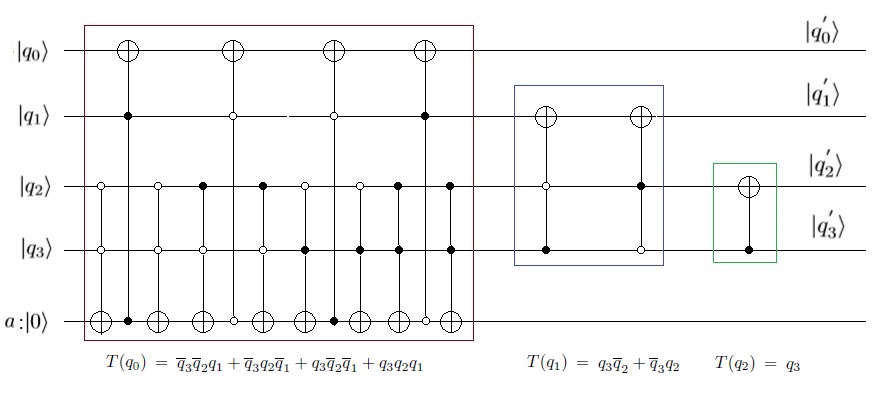}
	\caption{Implementation of the Quantum circuit using 2-input Toffoli and CNOT gates}
	\label{fig:Qcircuit2}
\end{figure}

The  quantum circuit is also simulated  using QISKIT  open-source framework simulator for quantum circuit \cite{QiskitURL,Qiskit}. In QISKIT, the Toffoli gate is composed of two control inputs and one output. To implement the circuit in Figure 	\ref{fig:Qcircuit1},  we redesigned the quantum circuit to include  2-input  Toffoli gates; however an additional  ancillary qubit is needed to store the  intermediate values. The design with 2-input Toffoli gates is presented in Figure \ref{fig:Qcircuit2}. The code to simulate the qunatum gray to binary converter is presented in Figure \ref{fig:QISKITCode}.

\pagebreak

The Q-Map algorithm can be summarised as follows:
\begin{itemize}
	\item[] \textbf{Step} 1: For each qubit $q_{i}$ build the toggle function $T(q_i)=q_{i} \oplus q_{i}^{'}$ as the logical Exclusive-OR of the present state $q_{i}$ and the final state $q_{i}^{'}$. 
	\item[] \textbf{Step }2 :Establish the Q-Map of the switching function  $T(q_i)$ as a function of the $n$ qubits $(q_0^{'}, q_1^{'}, \dots ,q_{i-1}^{'},q_i\dots ,q_{n-1})$ .
	\item[] \textbf{Step} 3: In each cell inscribe the value of $T(q_i)$ in the Q-Map.
	\item[] \textbf{Step} 4: Find the  expression of $T(q_i)$ in Sum-Of-Product form  using the Q-Map such that:\\
	1. Groups should be as large as possible\\
	2. Group Overlapping is not allowed
	
	\item[] \textbf{Step} 5: For each expression $T(q_i)$ use the \textbf{CNOT-NOT-Toffoli} bases to implement the corresponding quantum circuit.
	
\end{itemize}

\pagebreak

\begin{figure}[H]
	\centering
	\includegraphics[width=1.0\textwidth]{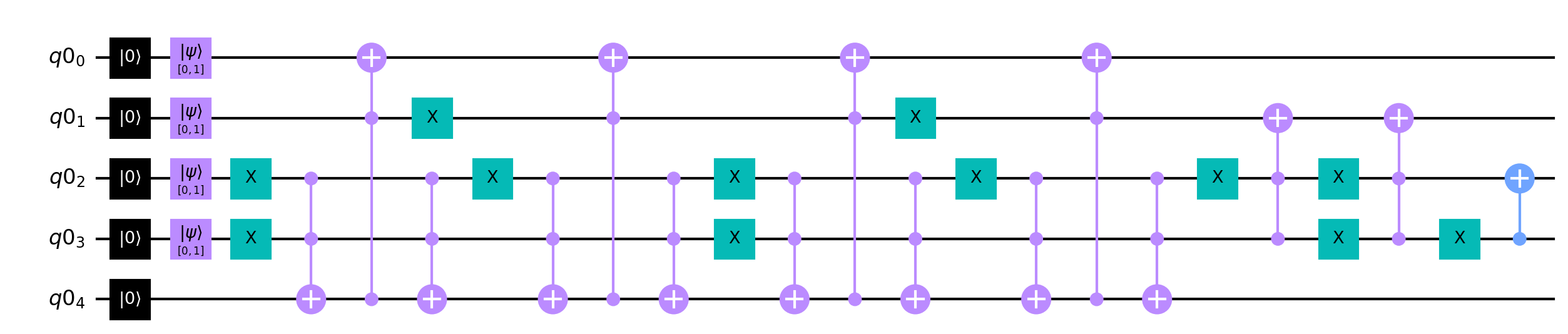}
	\caption{Implementation of the Quantum circuit using QISKIT}
	\label{fig:grayCodeQISKIT}
\end{figure}

\begin{figure}[H]
	\centering
	\begin{python}
		q = QuantumRegister(5)
		c = ClassicalRegister(5)
		grayEncoder = QuantumCircuit(q,c)
		# Reset input qbit q[0], q[1], q[2],q[3]
		grayEncoder.reset(q[0]) 
		grayEncoder.reset(q[1])
		grayEncoder.reset(q[2])
		grayEncoder.reset(q[3])
		#Reset ancillary qbit q[4]
		grayEncoder.reset(q[4])
		#========== calculate T(q0) =================
		grayEncoder.x(q[2]) # NOT gate
		grayEncoder.x(q[3]) # NOT gate
		grayEncoder.ccx(q[2],q[3],q[4]) # TOFOLLI gate
		grayEncoder.ccx(q[1],q[4],q[0]) 
		grayEncoder.ccx(q[2],q[3],q[4]) # TOFOLLI gate
		grayEncoder.x(q[2]) # NOT gate
		#----------------------------------------
		grayEncoder.ccx(q[2],q[3],q[4]) # TOFOLLI gate
		grayEncoder.x(q[1]) # not
		grayEncoder.ccx(q[1],q[4],q[0]) 
		grayEncoder.ccx(q[2],q[3],q[4]) # TOFOLLI gate
		grayEncoder.x(q[3]) # not
		#-------------------------------------
		grayEncoder.x(q[2]) # not
		grayEncoder.ccx(q[2],q[3],q[4]) # TOFOLLI gate
		grayEncoder.ccx(q[1],q[4],q[0]) 
		grayEncoder.x(q[1]) # not
		grayEncoder.ccx(q[2],q[3],q[4]) # TOFOLLI gate
		grayEncoder.x(q[2]) # not
		#--------------------
		grayEncoder.ccx(q[2],q[3],q[4]) # TOFOLLI gate
		grayEncoder.ccx(q[1],q[4],q[0]) 
		grayEncoder.ccx(q[2],q[3],q[4]) # TOFOLLI gate
		#=========== calculate T(q1) =================
		grayEncoder.x(q[2]) # not
		grayEncoder.ccx(q[2],q[3],q[1]) # TOFOLLI gate
		grayEncoder.x(q[2]) # not
		#---------------------------------
		grayEncoder.x(q[3]) # not
		grayEncoder.ccx(q[2],q[3],q[1]) # TOFOLLI gate
		grayEncoder.x(q[3]) # NOT gate
		#=========== calculate T(q2) =================
		grayEncoder.cx(q[3],q[2]) # CNOT gate
		
	\end{python}
	\caption{Code to implement  of the Quantum circuit in Fig. \ref{fig:grayCodeQISKIT}}
	\label{fig:QISKITCode}
\end{figure}

\section*{Q-Map optimisation} \label{secQmapOptimize}

	In classical boolean function minimsation using the Karnaugh map, the entry inscribed in every cell  indicates the value of the boolean function at the corresponding state. However in our proposed approach, the entry inscribed in each cell in the Q-map indicates change of state of the function. An entry of $1$ in th Q-map indicates that the function must change state (i.e toggle) and a value of $0$ indicates the function must remain in the same state. Thus including the $1's$ in the Q-map in an odd number of groups will result in one change of state and including the $0's$ in even number of groups will result in no change of state. This property of the Q-map could be used in our advantage to maximise the number of Q-map cells in each group  and produce a more simplified expression that  minimises the quantum cost of the design.\\

\begin{figure}[H]
	\centering
	\begin{subfigure}{.33\linewidth}
		\includegraphics[width=0.9\linewidth]{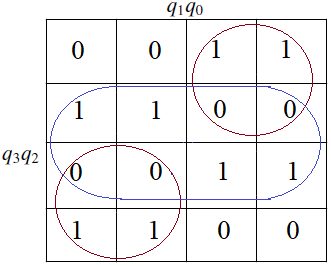}
		\caption{$T(q_{0})=f(q_0,q_1,q_2,q_3)$}
		\label{QmapTq0}
	\end{subfigure}
	\begin{subfigure}{.33\linewidth}
		\includegraphics[width=0.9\linewidth]{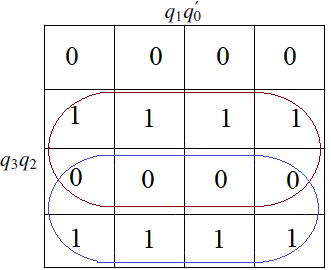}
		\caption{$T(q_{1})=f(q_0^{'},q_1,q_2,q_3)$}
		\label{QmapTq1}
	\end{subfigure}
	\medskip 
	
	\begin{subfigure}{.33\linewidth}
		\includegraphics[width=0.9\linewidth]{QMap_q2}
		\caption{$T(q_{2})=f(q_0^{'},q_1^{'},q_2,q_3)$}
		\label{QmapTq2}
	\end{subfigure}
	\begin{subfigure}{.33\linewidth}
		\includegraphics[width=0.9\linewidth]{QMap_q3}
		\caption{$T(q_{3})=f(q_0^{'},q_1^{'},q_2^{'},q_3)$}
		\label{QmapTq3}
	\end{subfigure}
	\caption{Optimised Q-Map representing the switching function for qubits $q_{0}$, $q_{1}$, $q_{2}$ and $q_{3}$,  $T(q_i)$. Every entry of $1$ must be included in an odd number of groups and an entry of $0$ must be included in an even number of groups.}
	\label{fig:QmapOpt}
\end{figure}

	The optimised Q-Map algorithm can be summarised as follows:
\begin{itemize}
	\item[] \textbf{Step} 1: For each qubit $q_{i}$ build the toggle function $T(q_i)=q_{i} \oplus q_{i}^{'}$ as the logical Exclusive-OR of the present state $q_{i}$ and the final state $q_{i}^{'}$. \\
	\item[] \textbf{Step }2 :Establish the Q-Map of the switching function  $T(q_i)$ as a function of the $n$ qubits $(q_0^{'}, q_1^{'}, \dots ,q_{i-1}^{'},q_i\dots ,q_{n-1})$ .\\
	\item[] \textbf{Step} 3: In each cell inscribe the value of $T(q_i)$ in the Q-Map.
	\item[] \textbf{Step} 4: Find the  expression of $T(q_i)$ in Sum-Of-Product form  using the Q-Map such that:
	\begin{enumerate}[itemsep=1pt,parsep=1pt]
		\item Groups should be as large as possible and may include $1's$ and $0's$.
		\item  Every $1$ must be included in an odd number of groups
		\item If a $0$ is included, it must be  included in an even number of groups
	\end{enumerate}
	\item[] \textbf{Step} 5: For each expression $T(q_i)$ use the \textbf{CNOT-NOT-Toffoli} bases to implement the corresponding quantum circuit.
	
\end{itemize}
	To demonstrate the idea, we re-evaluate the functions produces in the Q-map of Figure \ref{fig:Qmap} as shown in Figure \ref{fig:QmapOpt} and the functions  presented in Eq. \ref{eq:T} can be replaced by :

\begin{equation}\label{eq:Topt}
	\boxed{
		\begin{array}{rcl}
			T(q_{0}) &=& \overline{q}_{3} q_{1} +  q_{3} \overline{q}_{1} + q_{2}\\
			T(q_{1}) &=& q_{3}  +  q_{2}\\
			T(q_{2}) &=& q_{3}\\
			T(q_{3}) &=& 0
		\end{array}
	}
\end{equation}

Other alternative designs are possible, for example, $T(q_{1})$ can be also written as $\bar{q_2} + \bar{q_3}$.

\begin{figure}[H]
	\centering
	\includegraphics[width=0.8\linewidth]{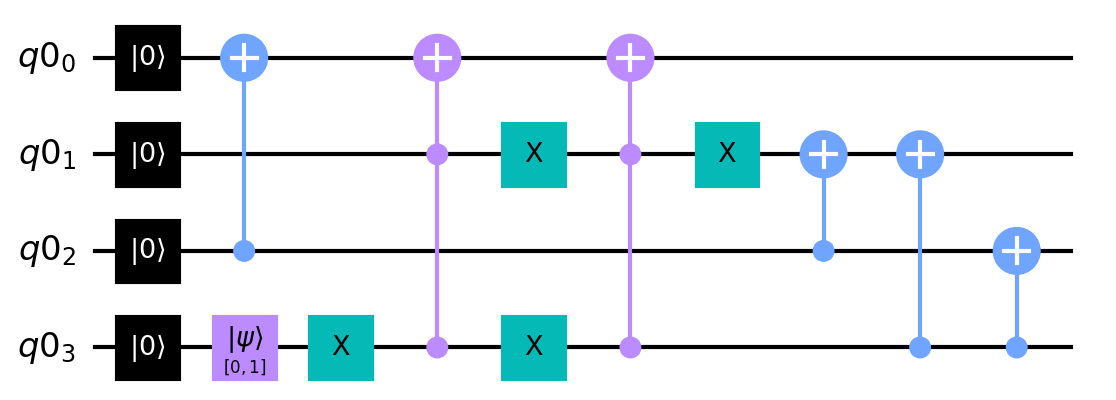}
	\centering \caption{ Implementation of the optimised Quantum circuit using QISKIT}
	\label{fig:circuitOpt}
\end{figure}

Based on the functions in Eq.\ref {eq:Topt}, the quantum circuit can be designed as  presented in  Figure \ref{fig:circuitOpt} and the corresponding QISKIT code is shown in Figure	\ref{fig:QISKITCodeopt}. The optimised approach re-designed the quantum circuit with no ancillary bits and significantly reduced the number of quantum gates.

\begin{figure}[h]
	\centering
	\begin{python}
		q = QuantumRegister(4)
		c = ClassicalRegister(4)
		grayEncoder = QuantumCircuit(q,c)
		
		# Reset input qbit q[0], q[1], q[2],q[3]
		grayEncoder.reset(q[0]) 
		grayEncoder.reset(q[1])
		grayEncoder.reset(q[2])
		grayEncoder.reset(q[3])
		
		#========== calculate T(q0) =================
		grayEncoder.cx(q[2],q[0]) 		# CNOT gate (T(q0) = q2)
		grayEncoder.x(q[3])       		# NOT gate
		grayEncoder.ccx(q[1],q[3],q[0]) # TOFOLLI gate (T(q0) = q1.!q3)
		grayEncoder.x(q[3]) 			# NOT gate
		grayEncoder.x(q[1]) 			# NOT gate
		grayEncoder.ccx(q[1],q[3],q[0]) # TOFOLLI gate T(q0) = !q1. q3
		grayEncoder.x(q[1]) 			# NOT gate
		
		#=========== calculate T(q1) =================
		grayEncoder.cx(q[2],q[1]) 		# TOFOLLI gate T(q1) = q2
		grayEncoder.cx(q[3],q[1]) 		# TOFOLLI gate T(q1) = q3
		
		#=========== calculate T(q2) =================
		grayEncoder.cx(q[3],q[2]) 		# CNOT gate	
	\end{python}
	\caption{Code to implement  of the Quantum circuit in Fig. \ref{fig:circuitOpt}}
	\label{fig:QISKITCodeopt}
\end{figure}

\section*{Conclusion}
In this paper we propose a visual method  for quantum realisation of classical  Boolean logic functions.  The proposed method utilise concepts from Boolean algebra to produce a quantum circuit with minimal number of quantum gates. The proposed technique is composed of three steps: (1) for each quantum bit $q_{i}$ build  a switching function $T(q_{i})$, (2) establish the Q-Map for $T(q_{i})$  and (3) using the Q-Map find the quantum expression to implement the switching function using the \textbf{NOT-CNOT-Toffoli} quantum gate basis.The  proposed method is demonstrated by implementing the Gray-Code Encoder.

\nolinenumbers

\bibliography{QmapBib}

\bibliographystyle{abbrv}

\end{document}